\documentclass[12pt] {article}
\usepackage{psfrag}
\usepackage{graphicx}
\usepackage{latexsym,amsfonts}
\usepackage{amssymb}
\begin{document}
\setcounter{page}{1}
\def\theequation{\arabic{section}.\arabic{equation}}
\def\theequation{\thesection.\arabic{equation}}
\setcounter{section}{0}

\title{Energy level displacement of  excited $np$ states of kaonic
  hydrogen}

\author{A. N. Ivanov\,\thanks{E--mail: ivanov@kph.tuwien.ac.at, Tel.:
+43--1--58801--14261, Fax: +43--1--58801--14299}~\thanks{Permanent
Address: State Polytechnic University, Department of Nuclear Physics,
195251 St. Petersburg, Russian Federation}\,,
M. Cargnelli\,\thanks{E--mail: michael.cargnelli@oeaw.ac.at}\,,
M. Faber\,\thanks{E--mail: faber@kph.tuwien.ac.at, Tel.:
+43--1--58801--14261, Fax: +43--1--58801--14299}\,, H.
Fuhrmann\,\thanks{E--mail: hermann.fuhrmann@oeaw.ac.at}\,,\\
V. A. Ivanova\,\thanks{E--mail: viola@kph.tuwien.ac.at, State
Polytechnic University, Department of Nuclear Physics, 195251
St. Petersburg, Russian Federation}\,, J. Marton\,\thanks{E--mail:
johann.marton@oeaw.ac.at}\,, N. I. Troitskaya\,\thanks{State
Polytechnic University, Department of Nuclear Physics, 195251
St. Petersburg, Russian Federation}~~, J. Zmeskal\,\thanks{E--mail:
johann.zmeskal@oeaw.ac.at}}
\date{\today}

\maketitle

\vspace{-0.5in}
\begin{center}
{\it Atominstitut der \"Osterreichischen Universit\"aten,
Arbeitsbereich Kernphysik und Nukleare Astrophysik, Technische
Universit\"at Wien, \\ Wiedner Hauptstr. 8-10, A-1040 Wien,
\"Osterreich \\ und\\ Stefan Meyer Institut f\"ur subatomare Physik 
der \"Osterreichische Akademie\\ der Wissenschaften,\\
Boltzmanngasse 3, A-1090, Wien, \"Osterreich}
\end{center}

\begin{abstract}
  We compute the energy level displacement of the excited $np$
  states of kaonic hydrogen within the quantum field theoretic and
  relativistic covariant model of strong low--energy $\bar{K}N$
  interactions suggested in EPJA {\bf 21}, 11 (2004). For the width of
  the energy level of the excited $2p$ state of kaonic hydrogen,
  caused by strong low--energy interactions, we find $\Gamma_{2p} =
  2\,{\rm meV} = 3\times 10^{12}\,{\rm sec^{-1}}$. This result is
  important for the theoretical analysis of the $X$--ray yields in
  kaonic hydrogen.\\ PACS: 36.10.Gv, 31.15.Ar, 32.30.Rj, 25.80.Nv
\end{abstract}

\maketitle

\newpage

\section{Introduction}

Recently \cite{IV3} we have computed the energy level displacement of
the ground state of kaonic hydrogen
\begin{eqnarray}\label{label1.1}
  - \epsilon^{(\rm th)}_{1s} + i\,\frac{\Gamma^{(\rm th)}_{1s}}{2} = 
(- 203 \pm 15) + i\,(113 \pm 14)\,{\rm eV}. 
\end{eqnarray} 
This result has been obtained within a quantum field theoretic and
relativistic covariant model of strong low--energy $\bar{K}N$
interactions near threshold of $K^-p$ scattering, based on the dominant
role of strange resonances $\Lambda(1405)$ and $\Sigma(1750)$ in the
s--channel of low--energy elastic and inelastic $K^-p$ scattering and
the exotic four--quark ( or $K\bar{K}$ molecules) scalar states
$a_0(980)$ and $f_0(980)$ in the $t$--channel of low--energy elastic
$K^- p$ scattering. 

The theoretical result (\ref{label1.1}) agrees well with recent
experimental data obtained by the DEAR Collaboration \cite{DEAR1}:
\begin{eqnarray}\label{label1.2}
  - \epsilon^{(\exp)}_{1s} + i\,\frac{\Gamma^{(\exp)}_{1s}}{2} = 
(- 194 \pm 41) + i\,(125 \pm 59)\,{\rm eV}. 
\end{eqnarray}
A systematic analysis of corrections, caused by electromagnetic and
QCD isospin--breaking interactions, to the energy level displacements
of the $ns$ states of kaonic hydrogen, where $n$ is the principal
quantum number, has been recently carried out by Mei\ss ner, Raha and
Rusetsky \cite{UM04} within Effective Field Theory by using the
non--relativistic effective Lagrangian approach based on Chiral
Perturbation Theory (ChPT) by Gasser and Leutwyler \cite{JG83,JG99}.
For the S--wave amplitude of $K^-N$ scattering near threshold,
computed in \cite{IV3,IV4}, the energy level displacement of the
ground state of kaonic hydrogen obtained by Mei\ss ner {\it et al.}
\cite{UM04} is equal to \[ - \epsilon^{(\rm th)}_{1s} +
i\,\frac{\Gamma^{(\rm th)}_{1s}}{2} = (- 266 \pm 23) + i\,(177 \pm
21)\,{\rm eV}.\] This agrees well with both our theoretical result
(\ref{label1.1}) and experimental data (\ref{label1.2}) within 1.5
standard deviations for the shift and one standard deviation for the
width.

In this paper we compute the energy level displacement of the excited
$np$ states of kaonic hydrogen, where $n$ is the principal quantum
  number and $p$ corresponds to the excited state with $\ell = 1$.
The knowledge of the energy level displacement of the excited $np$
states of kaonic hydrogen is very important for the understanding of
the accuracy of experimental measurements of the energy level
displacement of the ground state of kaonic hydrogen and the
theoretical analysis of the $X$--ray yields in kaonic hydrogen
\cite{DEAR1}, \cite{XR1}--\cite{XR7}.

The paper is organized as follows. In Section 2 we extend our approach
to the description of low--energy $K^-p$ interaction in the S--wave
state to the analysis of the low--energy $K^-p$ interaction in the
P--wave state with a total angular moment $J = 3/2$ and $J = 1/2$,
respectively.  We compute the P--wave scattering lengths of elastic
$K^-p$ scattering and the energy level shift of the $np$ excited state
of kaonic hydrogen. In Section 3 we compute the P--wave scattering
lengths of inelastic reactions $K^-p \to Y\pi$, where $Y\pi =
\Sigma^-\pi^+, \Sigma^+\pi^-, \Sigma^0\pi^0$ and $\Lambda^0\pi^0$. We
compute the energy level width of the $np$ excited state of kaonic
hydrogen. For the $2p$ state of kaonic hydrogen we get $\Gamma_{2p} =
2\,{\rm meV} = 3\times 10^{12}\,{\rm sec^{-1}}$. The  rate of
the hadronic decays of kaonic hydrogen from the $np$ excited state is
important for the theoretical analysis of the $X$--ray yields in 
kaonic hydrogen, which are the main experimental tool for the
measurement of the energy level displacement of the ground state of
kaonic hydrogen \cite{DEAR1}. In the Conclusion we discuss the
obtained results. 

\section{Energy level displacement of the $n\ell$ excited states of 
kaonic hydrogen. General formulas}
\setcounter{equation}{0}

According to \cite{IV2}, the energy level displacement of the excited
$n\ell$ states of kaonic hydrogen can be defined by 
\begin{eqnarray}\label{label2.1}
  \hspace{-0.5in}&& -\,\epsilon_{n\ell} + i\,\frac{\Gamma_{n\ell}}{2} = 
\frac{1}{2\ell +
    1}\sum^{\ell}_{m = -\ell}\int \frac{d^3k}{(2\pi)^3}
  \frac{\Phi^{\dagger}_{n \ell}(k)}{\sqrt{2 E_{K^-}(k)2 E_p(k)}}\int
  \frac{d^3q}{(2\pi)^3} \frac{\Phi_{n\ell}(q)}{\sqrt{2 E_{K^-}(q)2
      E_p(q)}}\nonumber\\
  \hspace{-0.5in}&&\times \int\!\!\!\int
  \frac{d\Omega_{\vec{k}}}{\sqrt{4\pi}}\frac{d\Omega_{\vec{q}}}{\sqrt{4\pi}}
\,Y^*_{\ell
    m}(\vartheta_{\vec{k}},\varphi_{\vec{k}})\,M(K^-(\vec{q}\,) 
p(-\vec{q},\sigma_p) \to 
  K^-(\vec{k}\,)p(-\vec{k},\sigma_p))\,Y_{\ell
    m}(\vartheta_{\vec{q}},\varphi_{\vec{q}}),
\end{eqnarray}
where $M(K^-(\vec{q}\,)p(-\vec{q},\sigma_p) \to
K^-(\vec{k}\,)p(-\vec{k},\sigma_p))$ is the amplitude of elastic $K^-
p$ scattering, $\Phi_{n\ell}(k)$ is a radial wave function of kaonic
hydrogen in the $n\ell$ excited state in momentum representation. It
is defined by \cite{IV2}
\begin{eqnarray}\label{label2.2}
\Phi_{n\ell}(k) =\sqrt{4\pi}
\int^{\infty}_0j_{\ell}(kr)R_{n\ell}(r)r^2 dr,
\end{eqnarray}
where $j_{\ell}(kr)$ are spherical Bessel functions \cite{MA72} and
$R_{n\ell}(r)$ is a radial wave function of kaonic hydrogen in the
coordinate representation \cite{LL65}:
\begin{eqnarray}\label{label2.3}
R_{n\ell}(r) = - \frac{2}{n^2}\,\sqrt{\frac{(n
- \ell - 1)!}{[(n + \ell)!]^3 a^3_B}}\,\Big(\frac{2 }{n }\frac{r}{
a_B}\Big)^{\ell}e^{\textstyle\,-r/n a_B}\,L^{2\ell + 1}_{n +
\ell}\Big(\frac{2 }{n }\frac{r}{ a_B}\Big).
\end{eqnarray}
Here $L^{2\ell + 1}_{n + \ell}(\rho)$ are the generalised Laguerre
polynomials given by \cite{LL65}
\begin{eqnarray}\label{label2.4}
L^{2\ell + 1}_{n + \ell}(\rho) = (-1)^{2\ell + 1}\,\frac{(n +
\ell)!}{(n - \ell - 1)!}\,\rho^{-(2\ell +
1)}\,e^{\textstyle\,\rho}\,\frac{d^{n - \ell - 1}}{d\rho^{n - \ell -
1}}(\rho^{n + \ell}e^{\textstyle\,-\rho}),
\end{eqnarray}
where $\rho = r/na_B$ and $a_B = 1/\alpha \mu = 83\,{\rm fm}$ is the
Bohr radius of kaonic hydrogen with $\mu = m_Km_N/(m_K + m_N) =
324\,{\rm MeV}$ and $\alpha = 1/137.036$ are the reduced mass of the
$K^-p$ pair, computed for $m_K = 494\,{\rm MeV}$ and $m_N = 940\,{\rm
  MeV}$, and the fine--structure constant, respectively. Spherical
harmonics $Y_{\ell m}(\vartheta,\varphi)$ are normalized by
\begin{eqnarray}\label{label2.5}
\int d\Omega\,Y^*_{\ell\,' m\,'}(\vartheta,\varphi)Y_{\ell
m}(\vartheta,\varphi) = \delta_{\ell\,'\ell}\,\delta_{m\,'m},
\end{eqnarray}
where $d\Omega = \sin\vartheta d\vartheta d\varphi$ is a volume
element of solid angle.

In Eq.(\ref{label2.1}) due to the wave functions $\Phi^*_{n\ell}(k)$
and $\Phi_{n\ell}(q)$ the integrand of the momentum integrals is
concentrated at momenta of order of $k \sim q \sim 1/na_B = \alpha
\mu/n = 2.4/n\,{\rm MeV}$. Therefore, the amplitude of elastic $K^-p$
scattering can be defined in the low--energy limit at $k, q \to 0$.
Since in the low--energy limit there is no spin--flip in the
transition $K^- + p \to K^- + p$ the amplitude of low--energy elastic
$K^-p$ scattering can be determined by \cite{SG67}--\cite{TE88} (see
also \cite{IV2}):
\begin{eqnarray}\label{label2.6}
  \hspace{-0.3in}&& M(K^-(\vec{q}\,)p(-\vec{q},\sigma_p) \to
  K^-(\vec{k}\,)p(-\vec{k},\sigma_p))=
  8\pi\,\sqrt{s}\sum^{\infty}_{\ell\,' = 0}[(\ell\,' +
  1)\,f_{\ell\,'+}(\sqrt{kq}) + \ell\,'\,f_{\ell\,'-}(\sqrt{kq})]
  \nonumber\\
  \hspace{-0.3in}&&\times\,P_{\ell\,'}(\cos\vartheta)=
  8\pi\,\sqrt{s}\sum^{\infty}_{\ell\,' = 0}[(\ell\,' +
  1)\,f_{\ell\,'+}(\sqrt{kq}) + \ell\,'\,f_{\ell\,'-}(\sqrt{kq})]\,
  \sum^{\ell\,'}_{m\,' = -\ell\,'}\frac{4\pi}{2 \ell\,' + 1}Y^*_{\ell\,'
    m\,'}(\vartheta_{\vec{q}},\varphi_{\vec{q}})\nonumber\\
  \hspace{-0.3in}&&\times\,Y_{\ell\,'
    m\,'}(\vartheta_{\vec{k}},\varphi_{\vec{k}}),
\end{eqnarray}
where $\sqrt{s}$ is the total energy in the $s$--channel of $K^-p$
scattering, $P_{\ell\,'}(\cos\vartheta)$ are Legendre polynomials
\cite{MA72} and $\vartheta$ is the angle between the relative momenta
$\vec{k}$ and $\vec{q}$. The amplitudes $f_{\ell\,'+}(\sqrt{kq})$ and
$f_{\ell\,'-}(\sqrt{kq})$ describe elastic $K^-p$ scattering in the
states with a total angular momentum $J = \ell\,' + 1/2$ and $J =
\ell\,' - 1/2$, respectively.  They are defined by
\begin{eqnarray}\label{label2.7}
  f_{\ell\,'+}(\sqrt{kq}) &=& \frac{1}{2i\sqrt{kq}}\,\Big(
  \eta_{\ell\,'+}(\sqrt{kq})e^{\textstyle\,+2 i\delta_{\ell\,'+}(\sqrt{kq})} - 
1\Big),\nonumber\\ 
f_{\ell\,'-}(\sqrt{kq}) &=&\frac{1}{2i\sqrt{kq}}\,\Big(
  \eta_{\ell\,'-}(\sqrt{kq})e^{\textstyle\,+2 i\delta_{\ell\,'-}(\sqrt{kq})} - 
1\Big),
\end{eqnarray}
where $\eta_{\ell\,'\pm}(\sqrt{kq})$ and
$\delta_{\ell\,'\pm}(\sqrt{kq})$ are inelasticities and phase shifts
of elastic $K^-p$ scattering \cite{SG67}--\cite{TE88}.

Near threshold the amplitudes $f_{\ell\,'+}(\sqrt{kq})$ and
$f_{\ell\,'-}(\sqrt{kq})$ possess the real and imaginary parts. The
real parts of the amplitudes $f_{\ell\,'+}(\sqrt{kq})$ and
$f_{\ell\,'-}(\sqrt{kq})$ are defined by the $\ell'$--wave scattering
lengths of $K^-p$ scattering \cite{MN79,IV2}
\begin{eqnarray}\label{label2.8}
{\cal R}e\,f_{\ell\,'+}(\sqrt{kq}) &=&
a^{K^-p}_{\ell\,'+}(kq)^{\,\ell\,'},\nonumber\\
{\cal Re}e\,f_{\ell\,'-}(\sqrt{kq}) &=&
a^{K^-p}_{\ell\,'-}(kq)^{\,\ell\,'}.
\end{eqnarray}
Using (\ref{label2.8}) for the shift of the energy level of the
$n\ell$ excited state of kaonic hydrogen we obtain \cite{IV2}
\begin{eqnarray}\label{label2.9}
\hspace{-0.3in}  \epsilon_{n\ell} = - \frac{2\pi}{\mu}\,\frac{(\ell +
    1)\,a^{K^-p}_{\ell+} +
    \ell\,a^{K^-p}_{\ell-}}{2\ell + 1}\,\Bigg|\int
  \frac{d^3k}{(2\pi)^3} \sqrt{\frac{m_K m_N}{
      E_{K^-}(k)E_p(k)}}\,k^{\,\ell}\,\Phi_{n \ell}(k)\Bigg|^2.
\end{eqnarray}
The imaginary parts of the amplitudes $f_{\ell\,'+}(\sqrt{kq})$ and
$f_{\ell\,'-}(\sqrt{kq})$ are defined by inelastic channels $K^- p \to
\Sigma^- \pi^+$, $K^- p \to \Sigma^+ \pi^- $, $K^- p \to \Sigma^0
\pi^0$ and $K^- p \to \Lambda^0 \pi^0$. According to \cite{IV2}, the
width $\Gamma_{n\ell}$ of the energy level of the $n\ell$ excited 
state of kaonic hydrogen is given by
\begin{eqnarray}\label{label2.10}
  \Gamma_{n\ell} &=&  \frac{4\pi}{\mu} \sum_{Y\pi}
  \Big[\frac{(\ell +
    1)\,a^{Y\pi}_{\ell+} +
    \ell\,a^{Y\pi}_{\ell-}}{2\ell + 1}\Big]^2\,
  [k_{Y\pi}(W_{n\ell})]^{\,2\ell +
    1}\nonumber\\
  &&\times\,\Bigg|\int
  \frac{d^3k}{(2\pi)^3}\,\sqrt{\frac{m_Km_N} {E_{K^-}(k)
      E_p(k)}}\,k^{\,\ell}\,\Phi_{n\ell}(k)\Bigg|^2,
\end{eqnarray}
where we sum over all $Y\pi$ pairs $Y\pi = \Sigma^+ \pi^-,
\Sigma^- \pi^+, \Sigma^0 \pi^0$ and $\Lambda^0\pi^0$; 
$k_{Y\pi}(W_{n\ell})$ is a relative momentum of the $Y\pi$ pair
\begin{eqnarray}\label{label2.11}
 k_{Y\pi}(W_{n\ell}) = \frac{\sqrt{(W^2_{n\ell} - (m_Y + m_{\pi})^2)
(W^2_{n\ell} - (m_Y - m_{\pi})^2)}}{2 W_{n\ell}}
\end{eqnarray}
with  $W_{n\ell} = m_K + m_N + E_{n\ell}$ and $E_{n\ell}$ is the
binding energy of kaonic hydrogen in the $n\ell$ excited state
\cite{TE88}. 

The analysis of experimental data obtained by the DEAR Collaboration
\cite{DEAR1} requires the knowledge of the energy level displacement
of the excited $np$ states. For $\ell = 1$ the formulas 
(\ref{label2.9}) and (\ref{label2.10}) read
\begin{eqnarray}\label{label2.12}
  \hspace{-0.3in}\epsilon_{np} &=& -
  \frac{2\pi}{3}\,\frac{1}{\mu}\,(2 a^{K^-p}_{3/2} +
  a^{K^-p}_{1/2})\,\Bigg|\int \frac{d^3k}{(2\pi)^3}
  \sqrt{\frac{m_K m_N}{ E_{K^-}(k)E_p(k)}}\,k\,\Phi_{n
    p}(k)\Bigg|^2,\nonumber\\ \hspace{-0.3in} \Gamma_{np} &=&
  \frac{4\pi}{9 }\,\frac{1}{\mu}\sum_{Y\pi}(2 a^{Y\pi}_{3/2} +
  a^{Y\pi}_{1/2})^2\, k^3_{Y\pi}\,\Bigg|\int \frac{d^3k}{(2\pi)^3}
  \sqrt{\frac{m_K m_N}{ E_{K^-}(k)E_p(k)}}\,k\,\Phi_{n
    p}(k)\Bigg|^2,
\end{eqnarray}
where $\Phi_{n p}(k)$ is the radial wave function of kaonic hydrogen
in the $np$ excited state in the momentum representation, the indices
$3/2$ and $1/2$ denote the P--wave amplitudes of the reactions $K^- p
\to K^-p $ and $K^- p \to Y \pi$ with total angular momentum $J = 3/2$
and $J = 1/2$, respectively \cite{SG67}--\cite{TE88}.

The momentum integral in the r.h.s. of (\ref{label2.12}) has been
computed in \cite{IV2}. Using this result the energy level
displacement of the $np$  excited states reads
\begin{eqnarray}\label{label2.13}
  \hspace{-0.3in}\epsilon_{np} &=& -\,\frac{2}{3}\,\frac{\alpha^5}{n^3}\,
\Big(1 - \frac{1}{n^2}\Big)\,
  \Big(\frac{m_Km_N}{m_K + m_N}\Big)^4\,
  (2\,a^{K^-p}_{3/2} +
  a^{K^-p}_{1/2}),\nonumber\\
  \hspace{-0.3in} \Gamma_{np} &=& \frac{4}{9}\,\frac{\alpha^5}{n^3}\,
\Big(1 - \frac{1}{n^2}\Big)\,
  \Big(\frac{m_Km_N}{m_K + m_N}\Big)^4
  \sum_{Y\pi}(2\,a^{Y\pi}_{3/2} +
  a^{Y\pi}_{1/2})^2\,k^3_{Y\pi}.
\end{eqnarray}
Thus, the problem of the calculation of the energy level displacement
of the $np$ excited states of kaonic hydrogen reduces to the problem of
the calculation of the P--wave scattering lengths $a^{K^-p}_{1/2}$ and
$a^{K^-p}_{3/2}$ of elastic $K^-p$ scattering and P--wave scattering
lengths $a^{Y\pi}_{1/2}$ and $a^{Y\pi}_{3/2}$ of inelastic reactions
$K^- p \to Y \pi$ with $Y\pi = \Sigma^+ \pi^-, \Sigma^- \pi^+,
\Sigma^0 \pi^0$ and $\Lambda^0\pi^0$.

\section{Model for low--energy $K^-p$ scattering in the P--wave state}
\setcounter{equation}{0}

For the description of the P--wave amplitude
of low--energy $K^-p$ scattering we follow  \cite{IV3,IV4} and assume that \\
\noindent (i) the amplitudes with total angular momentum $J = 1/2$ are
defined by the contributions of the elastic background and the octets
of baryon resonances with spin $1/2$ and positive parity such as
$(N(1440), \Lambda^0(1600), \Sigma(1660)) = B_1({\bf 8}) $
and $(N(1710), \Lambda^0(1810), \Sigma(1880)) = B_2({\bf
  8})$ and \\
\noindent (ii) the amplitudes with total angular momentum $J = 3/2$ are
defined by the contributions of the elastic background and the baryon
resonances with spin $3/2$ and positive parity from decuplet
$(\Delta(1232), \Sigma(1385)) = B_3({\bf 10})$ and octet
$(N(1720), \Lambda^0(1890), \Sigma(1840)) = B_4({\bf 8})$
\cite{PDG00}. We would like to emphasize that the baryon resonances we
will treat as elementary particles defined by local fields and local
phenomenological Lagrangians with phenomenological coupling constants
\cite{IV3,IV4}( see also \cite{MR96}). The contribution of
  the octet of low--lying baryons baryons with spin $1/2$ and positive
  parity $(N(940), \Lambda^0(1116), \Sigma(1193)) = B({\bf
    8})$ we include to the elastic background.

\subsection{P--wave scattering lengths of  elastic $K^- p$ scattering}

The P--wave amplitude of elastic $K^-p$ scattering at threshold is
defined by two P--wave scattering lengths $a^{K^-p}_{1/2}$ and
$a^{K^-p}_{3/2}$ caused by the interactions of the $K^-p$ pair in the
states with a total angular momentum $J = 1/2$ and $J = 3/2$, respectively.
 
\vspace{0.1in}
\noindent{\bf P--wave scattering length $a^{K^-p}_{1/2}$}
\vspace{0.1in}

According to our approach to the description of the low--energy $K^-p$
interaction in the S--wave state extended to the low--energy $K^-p$
interaction in the P--wave state, the amplitude
$a^{K^-p}_{1/2}$ has the following form 
\begin{eqnarray}\label{label3.1}
a^{K^-p}_{1/2} = (a^{K^-p}_{1/2})_B + \sum_R (a^{K^-p}_{1/2})_R\,, 
\end{eqnarray}
where $(a^{K^-p}_{1/2})_B$ is the contribution of an elastic background
and $(a^{K^-p}_{1/2})_R$ is the contribution of the baryon resonance $R
= \Lambda^0_1, \Sigma^0_1, \Lambda^0_2$ and $\Sigma^0_2$. 

\vspace{0.1in}
\noindent{\bf Resonance contribution to  P--wave scattering length
  $a^{K^-p}_{1/2}$}
\vspace{0.1in}

The phenomenological low--energy interactions $B_1({\bf 8})
B({\bf 8}) P({\bf 8})$ and $B_2({\bf 8})
B({\bf 8}) P({\bf 8})$, necessary for the calculation of
the contribution of the baryon resonances to the P--wave amplitude
$a^{K^-p}_{1/2}$, can be defined using the results obtained in
\cite{IV4}. As a result for the sum of the baryon resonance
contributions we obtain
\begin{eqnarray}\label{label3.2}
  \hspace{-0.3in} \sum_R (a^{K^-p}_{1/2})_R&=& 
  -\frac{1}{8\pi}\,
  \frac{1}{m_K + m_N}\,
  \Big[\frac{1}{\sqrt{3}} (3 - 2\alpha_1) g_{\pi NN_1}\Big]^2\,\frac{1}{2 m_N}\,
  \frac{1}{m_{\Lambda^0_1} - m_N - m_K}
  \nonumber\\
  \hspace{-0.3in}&&-\frac{1}{8\pi}\,
  \frac{1}{m_K + m_N}\,
  \Big[(2\alpha_1 - 1) g_{\pi NN_1}\Big]^2\,\frac{1}{2 m_N}\,
  \frac{1}{m_{\Sigma^0_1} - m_N - m_K}
\nonumber\\
 \hspace{-0.3in} &&-
  \frac{1}{8\pi}\,
  \frac{1}{m_K + m_N}\,
  \Big[\frac{1}{\sqrt{3}} (3 - 2\alpha_2) g_{\pi NN_2}\Big]^2\,\frac{1}{2 m_N}\,
  \frac{1}{m_{\Lambda^0_2} - m_N - m_K}
  \nonumber\\
  \hspace{-0.3in}&&-\frac{1}{8\pi}\,
  \frac{1}{m_K + m_N}\,
  \Big[(2\alpha_2 - 1) g_{\pi NN_2}\Big]^2\,\frac{1}{2 m_N}\,
  \frac{1}{m_{\Sigma^0_2} - m_N - m_K}.
\end{eqnarray}
The coupling constants of the interactions $K^-pB_1({\bf 8})$
and $K^-pB_2({\bf 8})$ are equal to: $g_{\pi NN_1} = 6.28$,
$\alpha_1 = 0.85$, $g_{\pi N N_2} = 1.20$ and $\alpha_2 = -
1.55$. These numerical values of the coupling constants one
  can obtain by using the phenomenological $SU(3)$--invariant
  interactions $B_1({\bf 8})B({\bf 8})P({\bf 8})$
  and $B_2({\bf 8})B({\bf 8})P({\bf 8})$ (see
  \cite{IV4}), where $P({\bf 8})$ is the octet of low--lying
  pseudoscalar mesons, and experimental data on the partial widths of
  the resonances $B_1({\bf 8})$ and $B_2({\bf 8})$
  \cite{PDG00}.  Using the recommended masses for the resonances
$m_{\Lambda^0_1} = 1600\,{\rm MeV}$, $m_{\Sigma_1} = 1660\,{\rm MeV}$,
$m_{\Lambda^0_2} = 1810\,{\rm MeV}$ and $m_{\Sigma_2} = 1880\,{\rm
  MeV}$ we compute
\begin{eqnarray}\label{label3.3}
 \sum_R (a^{K^-p}_{1/2})_R = -\,0.013\,{\rm m^{-3}_{\pi}}.
\end{eqnarray}
Now we proceed to computing the contribution of the elastic background
$(a^{K^-p}_{1/2})_B$.

\vspace{0.1in}
\noindent{\bf Elastic background contribution to the P--wave
  scattering length $a^{K^-p}_{1/2}$}
\vspace{0.1in}

According to \cite{IV3}, the contribution of the elastic background
$(a^{K^-p}_{1/2})_B$ to the P--wave scattering length $a^{K^-p}_{1/2}$
should be defined by the contribution of all low--energy interactions
$(a^{K^-p}_{1/2})_{CA}$, which can be described within the Effective
Chiral Lagrangian (the ECL) approach \cite{SG69} or that is equivalent
within Current Algebra (CA) \cite{SA68}--\cite{EK98}, supplemented by
soft--kaon theorems (SKT) \cite{ER72}--\cite{EK98}, and the
contribution $(a^{K^-p}_{1/2})_{K\bar{K}}$ of low--energy exchanges
with the exotic scalar mesons $a_0(980)$ and $f_0(980)$, which are
four--quark states \cite{RJ77,NA80} or $K\bar{K}$ molecules
\cite{NA80,SK03}. The description of strong low--energy interactions
of these mesons goes beyond the ECL approach, describing strong
low--energy interactions of mesons with $q\bar{q}$ and baryons with
$qqq$ quark structures.  Recent experimental confirmation of the
exotic structure of the scalar mesons $a_0(980)$ and $f_0(980)$ has
been obtained by the DEAR Collaboration at DAPHNE \cite{DAPHNE}.

Thus, the P--wave scattering
length $(a^{K^-p}_{1/2})_B$ is defined by
\begin{eqnarray}\label{label3.4}
 (a^{K^-p}_{1/2})_B = (a^{K^-p}_{1/2})_{CA} + (a^{K^-p}_{1/2})_{K\bar{K}}.
\end{eqnarray}
Using the results obtained in \cite{IV3} we compute the contribution
of the exotic scalar mesons 
\begin{eqnarray}\label{label3.5}
  a^{K^-p}_{1/2,K\bar{K}} = -\,\frac{1}{\pi}\,\frac{m_N}{m_K + m_N}\,
  \frac{g_{\,D} g_{\,0}}{~~m^4_{a_{\,0}}}\,\Big(1 - \frac{1}{8}\,
\frac{m^2_{a_{\,0}}}{m^2_N}\Big)
 = - \,0.018\,{\rm m^{-3}_{\pi}},
\end{eqnarray}
where $m_{f_{\,0}} = m_{a_{\,}0} = 980\,{\rm MeV}$, $g_{\,0} = g_{a_0 K^+K^-} =
g_{f_0 K^+K^-} = 2746\,{\rm MeV}$ \cite{NA80} and $g_{\,D} = \xi\,g_{\pi
  NN}/g_A = 0.95\,g_{\pi NN}$. For the calculation of $g_D$
  we have used $\xi = 1.2$ \cite{IV3} and $g_A = 1.267$ \cite{PDG00}.
  The coupling constant $g_{\pi NN}$ of the $\pi NN$ interaction is
  equal to $g_{\pi NN} = 13.21$ \cite{PSI2} (see also \cite{TE03} by
  Ericson, Loiseau and Wycech, where the authors have obtained
  $g_{\pi NN} = 13.28 \pm 0.08$).

The contribution to the P--wave amplitude, caused by the ECL
interactions, we represent in the form of the superposition of the
contributions of the $\Lambda^0(1116)$ and $\Sigma^0(1193)$ hyperon
exchanges and the term $(a^{K^-p}_{1/2})_{SKT}$, which can be
computed applying the soft--kaon technique \cite{ER72}--\cite{EK98}.
Thus, we get
\begin{eqnarray}\label{label3.6}
 (a^{K^-p}_{1/2})_{CA} &=& (a^{K^-p}_{1/2})_{SKT}\nonumber\\
 &-&
  \frac{1}{8\pi}\,
  \frac{1}{m_K + m_N}\,
  \Big[\frac{1}{\sqrt{3}} (3 - 2\alpha) g_{\pi NN}\Big]^2\,\frac{1}{2 m_N}\,
  \frac{1}{m_{\Lambda^0} - m_N - m_K}
  \nonumber\\
  &-&\frac{1}{8\pi}\,
  \frac{1}{m_K + m_N}\,
  \Big[(2\alpha - 1) g_{\pi NN}\Big]^2\,\frac{1}{2 m_N}\,
  \frac{1}{m_{\Sigma^0} - m_N + m_K}.
\end{eqnarray}
For $m_{\Lambda^0} = 1116\,{\rm MeV}$, $\Sigma^0 = 1193\,{\rm MeV}$,
$g_{\pi NN} = 13.21$ and $\alpha = 0.64$ \cite{IV4} we obtain
\begin{eqnarray}\label{label3.7}
 (a^{K^-p}_{1/2})_{CA} = (a^{K^-p}_{1/2})_{SKT}  + 0.024\,{\rm m^{-3}_{\pi}}.
\end{eqnarray}
Summing up the contributions, for the P--wave scatting length
$a^{K^-p}_{1/2}$ of $K^-p$ scattering with a total angular momentum $J = 1/2$
we get
\begin{eqnarray}\label{label3.8}
 a^{K^-p}_{1/2} = (a^{K^-p}_{1/2})_{SKT}  - 0.007\,{\rm m^{-3}_{\pi}}.
\end{eqnarray}
We suggest to compute the quantity $(a^{K^-p}_{1/2})_{SKT}$ together
with $(a^{K^-p}_{3/2})_{SKT}$, the contribution of the elastic
background to the P--wave scattering length $a^{K^-p}_{3/2}$ of $K^-p$
scattering with a total angular momentum $J = 3/2$.

\vspace{0.1in}
\noindent{\bf P--wave scattering length $a^{K^-p}_{3/2}$}
\vspace{0.1in}

The P--wave scattering length $a^{K^-p}_{3/2}$ we represent by
\begin{eqnarray}\label{label3.9}
a^{K^-p}_{3/2} = (a^{K^-p}_{3/2})_B + \sum_R (a^{K^-p}_{3/2})_R\,, 
\end{eqnarray}
where $(a^{K^-p}_{3/2})_B$ is the contribution of an elastic
background and $(a^{K^-p}_{3/2})_R$ is the contribution of the baryon
resonances $R = \Sigma^0_3, \Lambda^0_4$ and $\Sigma^0_4$. The
elastic background $(a^{K^-p}_{3/2})_B$ does not contain rapidly
changing contributions, therefore below we assume that
$(a^{K^-p}_{3/2})_B = (a^{K^-p}_{3/2})_{SKT}$.

\vspace{0.1in}
\noindent{\bf Resonance contribution to the P--wave scattering length
  $a^{K^-p}_{3/2}$}
\vspace{0.1in}

The phenomenological low--energy interaction of the resonance
$\Sigma^0_3$ with octets low--lying baryons $B({\bf 8})$ and
pseudoscalar mesons $P({\bf 8})$ is defined by
\cite{MN79,TE88,AI99} (see also \cite{PDG00}):
\begin{eqnarray}\label{label3.10} 
  \hspace{-0.3in}&&{\cal L}_{\Sigma^0_3BP}(x) =\nonumber\\
  \hspace{-0.3in}&&=\frac{g_{\pi NN}}{\sqrt{6} m_N}\,\bar{\Sigma}^0_{3\mu}(x)[ 
  \Sigma^+(x)\partial^{\mu} \pi^-(x) - \Sigma^-(x)\partial^{\mu} \pi^+(x) 
  + p(x)
  \partial^{\mu} K^-(x) + \sqrt{3}\,
  \Lambda^0(x)\partial^{\mu}\pi^0(x)]\nonumber\\
  \hspace{-0.3in}&&+ \frac{g_{\pi NN}}{\sqrt{6} m_N}\,[ 
  \bar{\Sigma}^+(x)\partial^{\mu} \pi^+(x) - \bar{\Sigma}^-(x)\partial^{\mu}
  \pi^-(x) - \bar{p}(x)
  \partial^{\mu} K^+(x) + \sqrt{3}\,
  \bar{\Lambda}^0(x)\partial^{\mu}\pi^0(x)]\,\Sigma^0_{3\mu}(x),\nonumber\\
  \hspace{-0.3in}&&
\end{eqnarray}
where we have written down only those interactions which contribution to
the P--wave amplitude of low--energy $K^-p$ scattering.

Using (\ref{label3.10}) we compute the contribution of the resonance
$\Sigma(1385)$ to the P--wave scattering length
$a^{K^-p}_{3/2}$:
\begin{eqnarray}\label{label3.11}
 \hspace{-0.3in} &&(a^{K^-p}_{3/2})_{\Sigma^0_3} = -\,\frac{g^2_{\pi NN}}{36 
\pi m_N}\,\frac{1}{m_K + m_N}\,
  \frac{m_{\Sigma^0_3}}{m^2_{\Sigma^0_3} - (m_K + m_N)^2}\, 
  \Big\{\Big[1 - \frac{1}{2}\,\frac{m_K}{m_N} - \frac{1}{4}\,
  \frac{m^2_K }{m^2_N}\Big]\nonumber\\
  \hspace{-0.3in}&&+ \frac{(m_K + m_N)}{m_{\Sigma^0_3}}\Big[1 + \frac{1}{2}\,
  \frac{m_K}{m_N}\,\frac{(m_K + m_N)}{m_{\Sigma^0_3}}- \frac{1}{4}\,
  \frac{m^2_K }{m^2_N}\,\Big(1 + \frac{(m_K + m_N)}{m_{\Sigma^0_3}} - \frac{(m_K + m_N)^2}{m^2_{\Sigma^0_3}}
  \Big)\Big] \Big\} =\nonumber\\ 
\hspace{-0.3in}&&= 0.060\,{\rm m^{-3}_{\pi}}.
\end{eqnarray}
The contribution of the resonances $\Lambda^0_4$ and $\Sigma^0_4$ to
$a^{K^-p}_{3/2}$ is equal to
\begin{eqnarray}\label{label3.12}
  \hspace{-0.3in}\sum_{R = \Lambda^0_4,\Sigma^0_4}(a^{K^-p}_{3/2})_R &=& 
  -\,\frac{1}{6 \pi m_N}\,\Big[\frac{1}{\sqrt{3}} (3 - 2\alpha_4) 
  g_{\pi NN_4}\Big]^2\,
  \frac{1}{m_K + m_N}\,\frac{m_{\Lambda^0_4}}{m^2_{\Lambda^0_4} - (m_K + m_N)^2}\nonumber\\\, 
  \hspace{-0.3in}&&\times\,\Big\{\Big[1 - \frac{1}{2}\,
  \frac{m_K}{m_N} - \frac{1}{4}\,
  \frac{m^2_K }{m^2_N}\Big] + \frac{(m_K + m_N)}{m_{\Lambda^0_4}}\,
\Big[1 + \frac{1}{2}\,
  \frac{m_K}{m_N}\,\frac{(m_K + m_N)}{m_{\Lambda^0_4}}\nonumber\\
  \hspace{-0.3in}&&- \frac{1}{4}\,
  \frac{m^2_K }{m^2_N}\,\Big(1 + \frac{(m_K + m_N)}{m_{\Lambda^0_4}} 
- \frac{(m_K + m_N)^2}{m^2_{\Lambda^0_4}}
  \Big)\Big] \Big\}\nonumber\\
  \hspace{-0.3in}&&-\frac{1}{6 \pi m_N}\,\,
  \Big[(2\alpha_4 - 1) g_{\pi NN_4}\Big]^2\,
  \frac{1}{m_K + m_N}\,\frac{m_{\Sigma^0_4}}{m^2_{\Sigma^0_4} - (m_K + m_N)^2}\nonumber\\\, 
  \hspace{-0.3in}&&\times\,\Big\{\Big[1 - \frac{1}{2}\,
  \frac{m_K}{m_N} - \frac{1}{4}\,
  \frac{m^2_K }{m^2_N}\Big] + \frac{(m_K + m_N)}{m_{\Sigma^0_4}}\,\Big[1 + \frac{1}{2}\,
  \frac{m_K}{m_N}\,\frac{(m_K + m_N)}{m_{\Sigma^0_4}}\nonumber\\
  \hspace{-0.3in}&&- \frac{1}{4}\,
  \frac{m^2_K }{m^2_N}\,\Big(1 + \frac{(m_K + m_N)}{m_{\Sigma^0_4}} 
- \frac{(m_K + m_N)^2}{m^2_{\Sigma^0_4}}
  \Big)\Big] \Big\}.
\end{eqnarray}
Using the experimental data on the resonances from the octet
$B_4({\bf 8})$ \cite{PDG00} we compute the coupling constants
$g_{\pi NN_4} = 1.16$ and $\alpha_4 = 0.32$. For $m_{\Lambda^0_4} =
1890\,{\rm MeV}$ and $m_{\Sigma^0_4} = 1840\,{\rm MeV}$ the numerical
value of the contribution of the resonances $\Lambda^0_4$ and
$\Sigma^0_4$ to the P--wave scattering length $a^{K^-p}_{3/2}$ reads
\begin{eqnarray}\label{label3.13}
  \sum_{R = \Lambda^0_4,\Sigma^0_4}(a^{K^-p}_{3/2})_R = -\, 0.001\,{\rm m^{-3}_{\pi}}.
\end{eqnarray}
The P--wave scattering length of $K^-p$ scattering with total angular momentum
$J = 3/2$ is given by
\begin{eqnarray}\label{label3.14}
 a^{K^-p}_{3/2} = (a^{K^-p}_{3/2})_{SKT} + 0.059\,{\rm m^{-3}_{\pi}}.
\end{eqnarray}
Summing up the contributions (\ref{label3.8}) and (\ref{label3.14}) we
obtain the total P--wave scattering length of elastic $K^-p$
scattering in the P--wave state
\begin{eqnarray}\label{label3.15}
  2\,a^{K^-p}_{3/2} + a^{K^-p}_{1/2} = (2\,a^{K^-p}_{3/2} +  a^{K^-p}_{1/2})_{SKT} 
+ 0.111\,{\rm m^{-3}_{\pi}}.
\end{eqnarray}
Now we turn to the calculation of the term $(2\,a^{K^-p}_{3/2}
+ a^{K^-p}_{1/2})_{SKT}$.

\subsection{Soft--kaon theorem for  amplitude of  elastic $K^-p$
  scattering and elastic P--wave background}

Soft--kaon theorems, as a part of ChPT \cite{JG83,JG99}, define
amplitudes of low--energy reactions with kaons as expansions in powers
of 4--momenta of kaons $k$, with kaons treated off--mass shell $k^2
\neq m^2_K$.  Using the reduction technique and the PCAC hypothesis
\cite{SA68}--\cite{EK98} the $S$--matrix element of elastic
low--energy transition $K^-p \to K^-p$ can be defined by
\begin{eqnarray}\label{label3.16}
 \hspace{-0.3in}&&\langle out; K^-(\vec{k}\,) p(-\vec{k},\sigma_p) | 
  K^-(\vec{q}\,) p(-\vec{q},\sigma_p); in\rangle  = -\, 
  \frac{(m^2_K  - k^2)}{\sqrt{2}\,F_K m^2_K }\, 
  \frac{(m^2_K  - q^2)}{\sqrt{2}\,F_K m^2_K }\nonumber\\
  \hspace{-0.3in}&&\times 
  \int d^4xd^4y\,e^{\textstyle\,+i k\cdot x - iq\cdot y}\,\langle p(-\vec{k},
\sigma_p) |{\rm T}(\partial^{\mu}J^{4+i5}_{5\mu}(x)\partial^{\nu}J^{4-i5}_{5\nu}(y))|
  p(-\vec{q},\sigma_p)\rangle,
\end{eqnarray}
where ${\rm T}$ is a time--ordering operator and $J^{4 +
  i5}_{5\mu}(x)$ and $J^{4 - i5}_{5\nu}(x)$ are axial--vector hadronic
currents with quantum numbers of the $K^-$ and $K^+$ mesons
\cite{SA68,HP73}; $F_K = 113\,{\rm MeV}$ is the PCAC constant of
charged $K$ mesons. For further reduction of the r.h.s. of
Eq.(\ref{label3.16}) we use the relation \cite{SA68}
\begin{eqnarray}\label{label3.17}
  \hspace{-0.3in} && {\rm T}(\partial^{\mu}J^{4+i5}_{5\mu}(x) 
\partial^{\nu}J^{4-i5}_{5\nu}(y)) =  
  \frac{\partial}{\partial x_{\mu}}\frac{\partial}{\partial y_{\nu}} 
{\rm T}(J^{4+i5}_{5\mu}(x)J^{4-i5}_{5\nu}(y))\nonumber\\
  \hspace{-0.3in} &&- \frac{1}{2}\,\frac{\partial}{\partial x_{\mu}} 
\Big\{\delta(x^0 - y^0)\,[J^{4-i5}_{50}(y),J^{4+i5}_{5\mu}(x)]\Big\} 
-  \frac{1}{2}\,\frac{\partial}{\partial y_{\nu}}\Big\{\delta(x^0 - y^0) 
\,[J^{4+i5}_{50}(x),J^{4-i5}_{5\nu}(y)]\Big\}\nonumber\\
  \hspace{-0.3in}  &&-  \frac{1}{2}\,\delta(x^0 - y^0) 
\,[J^{4-i5}_{50}(y),\partial^{\mu} J^{4+i5}_{5\mu}(x)] - 
 \frac{1}{2}\,\delta(x^0 - y^0)\, 
[J^{4+i5}_{50}(x),\partial^{\nu} J^{4-i5}_{5\nu}(y)].
\end{eqnarray}
Substituting (\ref{label3.17}) into (\ref{label3.16}) and making
integration by parts and dropping surface terms we arrive at the
expression
\begin{eqnarray}\label{label3.18}
  \hspace{-0.3in}&&\langle out; K^-(\vec{k}\,) p(-\vec{k},\sigma_p) | 
  K^-(\vec{q}\,) p(-\vec{q},\sigma_p); in\rangle  = -\, 
  \frac{(m^2_K  - k^2)}{\sqrt{2}\,F_K m^2_K }\, 
  \frac{(m^2_K  - q^2)}{\sqrt{2}\,F_K m^2_K }\nonumber\\
  \hspace{-0.3in}&&\times 
  \int d^4xd^4y\,e^{\textstyle\,+i k\cdot x - iq\cdot y}\,\Big\{k^{\mu} q^{\nu} \langle p(-\vec{k},\sigma_p) |{\rm T}(J^{4+i5}_{5\mu}(x)J^{4-i5}_{5\nu}(y))|
  p(-\vec{q},\sigma_p)\rangle
  \nonumber\\
  \hspace{-0.3in}&& + \frac{1}{2}\,i\,k^{\mu}\,\delta(x^0 - y^0)\,\langle p(-\vec{k},\sigma_p) |[J^{4-i5}_{50}(y),J^{4+i5}_{5\mu}(x)]|
  p(-\vec{q},\sigma_p)\rangle\nonumber\\
  \hspace{-0.3in}&&  - \frac{1}{2}\,i\,q^{\nu}\,\delta(x^0 - y^0) 
  \,\langle p(-\vec{k},\sigma_p) |[J^{4+i5}_{50}(x),J^{4-i5}_{5\nu}(y)]|
  p(-\vec{q},\sigma_p)\rangle\nonumber\\
  \hspace{-0.3in}&&  -  \frac{1}{2}\,\delta(x^0 - y^0) 
  \,\langle p(-\vec{k},\sigma_p) |[J^{4-i5}_{50}(y),
  \partial^{\mu} J^{4+i5}_{5\mu}(x)]|
  p(-\vec{q},\sigma_p)\rangle\nonumber\\
  \hspace{-0.3in}&& - 
  \frac{1}{2}\,\delta(x^0 - y^0)\, 
  \langle p(-\vec{k},\sigma_p) |[J^{4+i5}_{50}(x), 
\partial^{\nu} J^{4-i5}_{5\nu}(y)]| p(-\vec{q},\sigma_p)\rangle\Big\}
\end{eqnarray}
From (\ref{label3.18}) we obtain the amplitude of elastic low--energy
$K^-p$ scattering with $K^-$ mesons off--mass shell. It reads
\begin{eqnarray}\label{label3.19}
  \hspace{-0.3in}&&M( K^-(\vec{q}\,) p(-\vec{q},\sigma_p) \to 
  K^-(\vec{k}\,) p(-\vec{k},\sigma_p)) =  
  \frac{(m^2_K  - k^2)}{\sqrt{2}\,F_K m^2_K }\, 
  \frac{(m^2_K  - q^2)}{\sqrt{2}\,F_K m^2_K }\nonumber\\
  \hspace{-0.3in}&&\times 
  i\int d^4x\,e^{\textstyle\,+i k\cdot x }\,\Big\{k^{\mu} q^{\nu} 
\langle p(-\vec{k},\sigma_p) |{\rm T}(J^{4+i5}_{5\mu}(x)J^{4-i5}_{5\nu}(0))|
  p(-\vec{q},\sigma_p)\rangle
  \nonumber\\
  \hspace{-0.3in}&& + \frac{1}{2}\,i\,k^{\mu}\,\delta(x^0)\,
\langle p(-\vec{k},\sigma_p) |[J^{4-i5}_{50}(0),J^{4+i5}_{5\mu}(x)]|
  p(-\vec{q},\sigma_p)\rangle\nonumber\\
  \hspace{-0.3in}&&  - \frac{1}{2}\,i\,q^{\nu}\,\delta(x^0) 
  \,\langle p(-\vec{k},\sigma_p) |[J^{4+i5}_{50}(x),J^{4-i5}_{5\nu}(0)]|
  p(-\vec{q},\sigma_p)\rangle\nonumber\\
  \hspace{-0.3in}&&  -  \frac{1}{2}\,\delta(x^0 ) 
  \,\langle p(-\vec{k},\sigma_p) |[J^{4-i5}_{50}(0),
  \partial^{\mu} J^{4+i5}_{5\mu}(x)]|
  p(-\vec{q},\sigma_p)\rangle\nonumber\\
  \hspace{-0.3in}&& - \frac{1}{2}\,\delta(x^0)\, 
  \langle p(-\vec{k},\sigma_p) 
  |[J^{4+i5}_{50}(x),\partial^{\nu} J^{4-i5}_{5\nu}(0)]|
  p(-\vec{q},\sigma_p)\rangle\Big\}.
\end{eqnarray}
The equal--time commutators read
\cite{SA68,HP73}
\begin{eqnarray}\label{label3.20}
  \delta(x^0)\,[J^{4+i5}_{50}(x),J^{4-i5}_{5\nu}(0)] &=& 
  (J^3_{\nu}(0) + \sqrt{3}\,J^8_{\nu}(0))\,\delta^{(4)}(x),\nonumber\\
  \delta(x^0)\,[J^{4+i5}_{50}(x),\partial^{\nu} J^{4-i5}_{5\nu}(0)] &=&
 -\,i\,(\sigma_{44}(0) + \sigma_{55}(0))\,\delta^{(4)}(x),
\end{eqnarray}
where $J^3_{\nu}(0)$ and $J^8_{\nu}(0)$ are vector hadronic currents,
related to the electromagnetic $J^{(\rm e m)}_{\nu}(0)$ and
hypercharge $Y_{\nu}(0)$ currents by 
\begin{eqnarray}\label{label3.21}
 J^3_{\nu}(0) + \sqrt{3}\,J^8_{\nu}(0) = J^{(\rm e m)}_{nu}(0) + Y_{\nu}(0),
\end{eqnarray}
and $\sigma_{ab}(0)$ is so--called {\it $\sigma$--term}
operator. The {\it $\sigma$--term} operator $\sigma_{ab}(0)$
  is related to the breaking of chiral symmetry. It can also be
  defined by the double commutator \cite{ER72}: $\sigma_{ab}(0) =
  [Q^a_5(0),[Q^b_5(0),H_{\chi SB}(0)]]$, where $Q^a_5(0)$ is the
  axial--vector charge operator and $H_{\chi SB}$ is the Hamiltonian
  of strong interactions breaking of chiral symmetry. In terms of
  current quark fields it reads $H_{\chi SB}(0) = m_u\,\bar{u}(0)u(0)
  + m_d\,\bar{d}(0)d(0)+ m_s\,\bar{s}(0)s(0)$, where $m_q\,(q =
  u,d,s)$ and $q(0) = u(0),d(0),s(0)$ are masses and interpolating
  fields of current quarks.

Substituting (\ref{label3.20}) into (\ref{label3.19}) and using
(\ref{label3.21}) we get 
\begin{eqnarray}\label{label3.22}
  \hspace{-0.3in}&&M( K^-(\vec{q}\,) p(-\vec{q},\sigma_p) \to 
  K^-(\vec{k}\,) p(-\vec{k},\sigma_p)) =  
  \frac{(m^2_K  - k^2)}{\sqrt{2}\,F_K m^2_K }\, 
  \frac{(m^2_K  - q^2)}{\sqrt{2}\,F_K m^2_K }\nonumber\\
  \hspace{-0.3in}&&\times \Big\{\,k^{\mu} q^{\nu} 
  \,i\int d^4x\,e^{\textstyle\,+i k\cdot x }\,\langle p(-\vec{k},\sigma_p)
  |{\rm T}(J^{4+i5}_{5\mu}(x)J^{4-i5}_{5\nu}(0))|
  p(-\vec{q},\sigma_p)\rangle
  \nonumber\\
  \hspace{-0.3in}&& + \frac{1}{2}\,(k^{\mu} + q^{\mu})\, 
  \langle p(-\vec{k},\sigma_p) |J^{(\rm em)}_{\mu}(0) + Y_{\mu}(0)|
  p(-\vec{q},\sigma_p)\rangle\nonumber\\
  \hspace{-0.3in}&&  - \,\langle p(-\vec{k},\sigma_p) |\sigma_{44}(0) + 
\sigma_{55}(0)| 
  p(-\vec{q},\sigma_p)\rangle \Big\}.
\end{eqnarray} 
The matrix elements of the {\it $\sigma$--term} operator can be
represented by \cite{MS77}--\cite{EK98,ST1}
\begin{eqnarray}\label{label3.23}
\hspace{-0.3in} &&\langle p(-\vec{k},\sigma_p) |\sigma_{44}(0) + 
\sigma_{55}(0)| 
  p(-\vec{q},\sigma_p)\rangle = 2\,\sigma^{(I = 1)}_{KN}(t)\,
\bar{u}(-\vec{k},\sigma_p)u(-\vec{q},\sigma_p),
\end{eqnarray}
where $\sigma^{(I = 1)}_{\bar{K}N}(t)$ is the scalar form factor
\cite{ER72}--\cite{EK98,ST1}, defining the contribution to the amplitude of
$\bar{K}N$ scattering in the state with isospin $I = 1$, and $t =
-\,(\vec{k} - \vec{q}\,)^2$ is a squared transferred momentum. In terms
of the quark--field operators the $\sigma$--term $\sigma^{(I =
  1)}_{\bar{K}N}(t) $ is defined by \cite{ER72,MS77,EK98,ST1}
\begin{eqnarray}\label{label3.24}
  \hspace{-0.3in} &&\sigma^{(I = 1)}_{KN}(t) = \frac{m_u + m_s}{4m_N}\,
\langle p(- \vec{k},\sigma_p)|\bar{u}(0)u(0) + \bar{s}(0)s(0)|p(- \vec{q},
\sigma_p)\rangle.
\end{eqnarray}
According to ChPT \cite{JG83,JG99}, the $\sigma$--term is of order of
squared 4--momenta of $K^-$--mesons, i.e. $\sigma^{(I = 1)}_{KN}(t)
\sim k^2 \sim q^2$.

Accounting for the contribution of the $K^-$--meson pole and keeping
the terms of order of $O(k^2)$ and $O(q^2)$ inclusively, we get the
following expression for the amplitude of elastic low--energy $K^-p$
scattering \cite{EK98}
\begin{eqnarray}\label{label3.25}
 \hspace{-0.3in} &&M( K^-(\vec{q}\,) p(-\vec{q},\sigma_p) \to 
  K^-(\vec{k}\,) p(-\vec{k},\sigma_p)) =\nonumber\\ 
 \hspace{-0.3in}  &&=\bar{u}(-\vec{k},\sigma_p)\Big\{
  \frac{F^p_E(t) + F^p_Y(t)}{4 F^2_K}\,(k + q)^{\mu}\gamma_{\mu} - 
 \frac{1}{F^2_K}\,\Big[\sigma^{(I = 1)}_{KN}(t) 
  - k^{\mu} q^{\nu}\,W_{\mu\nu}(\vec{k},\vec{q}\,)\Big]\Big\}
 u(-\vec{q},\sigma_p),\nonumber\\ 
 \hspace{-0.3in}  &&
\end{eqnarray} 
where we have denoted
\begin{eqnarray}\label{label3.26}
\hspace{-0.3in}&&\langle p(-\vec{k},\sigma_p) |J^{(\rm em)}_{\mu}(0) + Y_{\mu}(0)|
  p(-\vec{q},\sigma_p)\rangle = (F^p_E(t) + F^p_Y(t))\,\bar{u}(-\vec{k},\sigma_p)
\gamma_{\mu}u(-\vec{q},\sigma_p),\nonumber\\
\hspace{-0.3in}&&\frac{1}{2}\,i\int d^4x\,\langle p(-\vec{k},\sigma_p)
  |{\rm T}(J^{4+i5}_{5\mu}(x)J^{4-i5}_{5\nu}(0))|
  p(-\vec{q},\sigma_p)\rangle = \nonumber\\
\hspace{-0.3in}&&= \bar{u}(-\vec{k},\sigma_p)\,W_{\mu\nu}(\vec{k},\vec{q}\,)
\,u(-\vec{q},\sigma_p).
\end{eqnarray} 
Here $F^p_E(t)$ and $F^p_Y(t)$ are the form factors of the electric and
hypercharge of the proton, normalized by $F^p_E(0) = F^p_Y(0) =
1$. We have not taken into account the magnetic form factor,
  which does not contribute to the S-- and P--wave amplitudes of
  $K^-p$ scattering at threshold.

The last two terms in Eq.(\ref{label3.25}) are of order of $O(k^2)$,
where $k^2 \sim q^2 \sim k\cdot q$. For the calculation of the P--wave
scattering length of elastic $K^-p$ scattering the contribution of the
terms of order of $O(k^2)$ can be neglected. 

From Eq.(\ref{label3.25}) at leading order in chiral expansion
\cite{JG83,JG99} we obtain the contribution to the P--wave amplitude
of low--energy elastic $K^-p$ scattering 
\begin{eqnarray}\label{label3.27}
 (2\,a^{K^-p}_{3/2} +  a^{K^-p}_{1/2})_{SKT} = \frac{1}{16\pi}\,\frac{\mu}{F^2_K
}\,\frac{1}{m^2_N} = 
0.002\,{\rm m^{-3}_{\pi}}.
\end{eqnarray} 
Hence, the P--wave scattering length $(2\,a^{K^-p}_{3/2} +
a^{K^-p}_{1/2})_{SKT}$ is smaller than the contribution of the
resonance states and practically can be neglected for the calculation
of the P--wave scattering lengths of elastic $K^-p$ scattering and,
correspondingly, for the calculation of the energy level shift of the
$np$ excited state of kaonic hydrogen.  This implies that the P--wave
scattering lengths $(2\,a^{Y\pi}_{3/2} + a^{Y\pi}_{1/2})_{SKT}$ can
also be neglected in comparison with the contributions of the
resonance states.

\subsection{P--wave scattering length $2 a^{K^-p}_{3/2} +
  a^{K^-p}_{1/2}$ of elastic $K^-p$ scattering and energy level shift
  of $np$ excited state of kaonic hydrogen}

Substituting (\ref{label3.27}) into (\ref{label3.15}) we obtain the
P--wave scattering length of elastic $K^-p$ scattering
\begin{eqnarray}\label{label3.29}
 2\,a^{K^-p}_{3/2} +  a^{K^-p}_{1/2} =
0.113\,{\rm m^{-3}_{\pi}}.
\end{eqnarray} 
Using (\ref{label3.29}) we compute the shift of the energy level of the 
$np$ excited state of kaonic hydrogen, given by Eq.(\ref{label2.13}). We get
\begin{eqnarray}\label{label3.30}
  \epsilon_{np}  = \frac{32}{3}\,\frac{1}{n^3}\,
\Big(1 - \frac{1}{n^2}\Big)\,\epsilon_{2p},
\end{eqnarray}
where the shift of the energy level of the $2p$ excited state is equal to
\begin{eqnarray}\label{label3.31}
  \epsilon_{np}  = -\,\frac{\alpha^5}{16}\,\Big(\frac{m_Km_N}{m_K + m_N}\Big)^4\,
  (2\,a^{K^-p}_{3/2} +
  a^{K^-p}_{1/2}) = -\,0.6\,{\rm meV}. 
\end{eqnarray}
Hence, the shift of the energy level $\epsilon_{np}$ of the $np$
excited state of kaonic hydrogen, induced by strong low--energy
interactions, is smaller than $1\,{\rm meV}$,  i.e. $|\epsilon_{np}| <
1\,{\rm meV}$. 

We would like to emphasize that unlike the shift of the energy level
of the $ns$ state of kaonic hydrogen, which is defined by repulsive
forces $\epsilon_{ns} = (203 \pm 15)/n^3\,{\rm eV}$ \cite{IV3}, the
shift of the energy level of the $np$ excited state $\epsilon_{np}$,
given by Eq.(\ref{label3.30}), is caused by  attractive forces.

\section{P--wave scattering lengths $2 a^{Y\pi}_{3/2} +
  a^{Y\pi}_{1/2}$ of inelastic channels $K^-p \to Y\pi$}
\setcounter{equation}{0}

The imaginary part of the P--wave amplitude of elastic $K^-p$
scattering at threshold, defining the total width of the excited $np$
state of kaonic hydrogen, is caused by the four opened inelastic
channels $K^-p \to \Sigma^+\pi^-$, $K^-p \to \Sigma^-\pi^+$, $K^-p \to
\Sigma^0\pi^0$ and $K^-p \to \Lambda^0\pi^0$. At threshold the
contribution of these inelastic channels we describe by the P--wave
scattering lengths $a^{Y\pi}_{1/2}$ and $a^{Y\pi}_{3/2}$ with $Y\pi =
\Sigma^+\pi^-$, $\Sigma^-\pi^+$, $\Sigma^0\pi^0$ and $\Lambda^0\pi^0$,
respectively. The P--wave scattering lengths $a^{Y\pi}_{1/2}$ and
$a^{Y\pi}_{3/2}$ determine low-energy transitions $K^-p\to Y\pi$ with
total angular moment $J =1/2$ and $J = 3/2$, respectively.

The P--wave scattering lengths $a^{Y\pi}_J$ we represent in the form
of the superposition of the background part $(a^{Y\pi}_J)_B$ and the
resonant part $\sum_R(a^{Y\pi}_J)_R$. It is convenient to include the
contribution of the octet of low--lying baryons $B({\bf 8})
= (N(940), \Lambda^0(1116), \Sigma(1193))$ to the resonant part and to
define the contribution of the background as $(a^{Y\pi}_J)_B =
(a^{Y\pi}_J)_{SKT}$. Since, as has been shown above, the
  contribution of the resonances $\Lambda^0(1890)$ and
  $\Sigma^0(1840)$ is negligible small relative to the contribution of
  the resonance $\Sigma^0(1385)$, below for the calculation of the
  P--wave scattering lengths of inelastic channels $K^-p \to Y\pi$ we
  do not take them into account.

\subsection{P--wave scattering lengths $2 a^{\Sigma^+\pi^-}_{3/2} +
  a^{\Sigma^+\pi^-}_{1/2}$of inelastic channel $K^-p \to
  \Sigma^+\pi^-$}

The resonant parts of the  P--wave scattering lengths
$a^{\Sigma^+\pi^-}_{1/2}$ and $a^{\Sigma^+\pi^-}_{3/2}$ of the
reaction $K^-p \to \Sigma^+\pi^-$ are equal to
\begin{eqnarray}\label{label4.1}
  && \sum_R(a^{\Sigma^+\pi^-}_{1/2})_R =\nonumber\\
  &&=\frac{1}{8\pi} \,\frac{1}{m_K + m_N}\, 
  \Big[
  \frac{1}{\sqrt{3}}\, (3 - 2\alpha)\, g_{\pi NN}\Big] 
  \Big[\frac{2}{\sqrt{3}}\, \alpha\, g_{\pi NN}\Big] 
  \frac{1}{2\sqrt{m_{\Sigma}m_N}} \,\frac{1}{m_{\Lambda^0} - m_K - m_N}
  \nonumber\\
  &&+ \frac{1}{8\pi} \,\frac{1}{m_K + m_N}\,\Big[
  \frac{1}{\sqrt{3}}\, (3 - 2\alpha_1)\, g_{\pi NN_1}\Big] 
  \Big[\frac{2}{\sqrt{3}}\, \alpha_1 \,g_{\pi NN_1}\Big] 
  \frac{1}{2\sqrt{m_{\Sigma}m_N}}\, \frac{1}{m_{\Lambda^0_1} - m_K - m_N}
  \nonumber\\
  &&+ \frac{1}{8\pi}\, \frac{1}{m_K + m_N}\,\Big[
  \frac{1}{\sqrt{3}}\, (3 - 2\alpha_2)\, g_{\pi NN_2}\Big] 
  \Big[\frac{2}{\sqrt{3}}\, \alpha_2\, g_{\pi NN_2}\Big] 
  \frac{1}{2\sqrt{m_{\Sigma}m_N}}\, \frac{1}{m_{\Lambda^0_2} - m_K - m_N}
  \nonumber\\
  &&+\frac{1}{8\pi}\, \frac{1}{m_K + m_N} 
  \Big[(2\alpha - 1)\, g_{\pi NN}\Big] 
  \Big[ 2\,(1 - \alpha)\, g_{\pi NN}\Big] 
  \frac{1}{2\sqrt{m_{\Sigma}m_N}}\, \frac{1}{m_{\Sigma^0} - m_K - m_N}
  \nonumber\\
  &&+ \frac{1}{8\pi}\, \frac{1}{m_K + m_N} 
  \Big[(2\alpha_1 - 1)\, g_{\pi NN_1}\Big] 
  \Big[ 2\, (1 - \alpha_1)\, g_{\pi NN_1}\Big] 
  \frac{1}{2\sqrt{m_{\Sigma}m_N}}\, \frac{1}{m_{\Sigma^0_1} - m_K - m_N}
  \nonumber\\
  &&+\frac{1}{8\pi} \,\frac{1}{m_K + m_N} 
  \Big[(2\alpha_2 - 1)\, g_{\pi NN_2}\Big] 
  \Big[ 2\, (1 - \alpha_2)\, g_{\pi NN_2}\Big] 
  \frac{1}{2\sqrt{m_{\Sigma}m_N}}\, \frac{1}{m_{\Sigma^0_2} - m_K - m_N}
  \nonumber\\
  && = (- 0.015 + 0.006 - 0.001 - 0.005 + 0.001 - 0.002)\,{\rm m^{-3}_{\pi}} = 
- 0.016\,{\rm m^{-3}_{\pi}}.
\end{eqnarray}
and 
\begin{eqnarray}\label{label4.2}
  (a^{\Sigma^+\pi^-}_{3/2})_R &=& \frac{g^2_{\pi NN}}{36\pi m_N}\,
  \frac{1}{m_K + m_N}\,\frac{1}{m_{\Sigma^0_3} - m_N - m_K}\,\sqrt{\frac{m_{\Sigma}}{m_N}}\,\Big(1 + \frac{1}{4}\,\frac{m_K}{m_N}\,\frac{m_K + m_N}{m_{\Sigma^0_3}}\Big) =\nonumber\\
&=& -\,0.082\,{\rm m^{-3}_{\pi}}.
\end{eqnarray}
The total P--wave scattering length of the reaction $K^-p \to
\Sigma^+\pi^-$ is equal to
\begin{eqnarray}\label{label4.3}
  2\,a^{\Sigma^+\pi^-}_{3/2} + a^{\Sigma^+\pi^-}_{1/2} = 
(2\,a^{\Sigma^+\pi^-}_{3/2} + a^{\Sigma^+\pi^-}_{1/2})_{SKT} 
- 0.180\,{\rm m^{-3}_{\pi}}.
\end{eqnarray}

\subsection{P--wave scattering lengths of $2 a^{\Sigma^-\pi^+}_{3/2} + 
  a^{\Sigma^-\pi^+}_{1/2}$of inelastic channel $K^-p \to
  \Sigma^-\pi^+$}

The resonant parts of the  P--wave scattering lengths
$a^{\Sigma^-\pi^+}_{1/2}$ and $a^{\Sigma^-\pi^+}_{3/2}$ of the
reaction $K^-p \to \Sigma^-\pi^+$ are equal to
\begin{eqnarray}\label{label4.4}
  && \sum_R(a^{\Sigma^-\pi^+}_{1/2})_R =\nonumber\\
  &&=\frac{1}{8\pi} \,\frac{1}{m_K + m_N}\, 
  \Big[
  \frac{1}{\sqrt{3}}\, (3 - 2\alpha)\, g_{\pi NN}\Big] 
  \Big[\frac{2}{\sqrt{3}}\, \alpha\, g_{\pi NN}\Big] 
  \frac{1}{2\sqrt{m_{\Sigma}m_N}} \,\frac{1}{m_{\Lambda^0} - m_K - m_N}
  \nonumber\\
  &&+ \frac{1}{8\pi} \,\frac{1}{m_K + m_N}\,\Big[
  \frac{1}{\sqrt{3}}\, (3 - 2\alpha_1)\, g_{\pi NN_1}\Big] 
  \Big[\frac{2}{\sqrt{3}}\, \alpha_1 \,g_{\pi NN_1}\Big] 
  \frac{1}{2\sqrt{m_{\Sigma}m_N}}\, \frac{1}{m_{\Lambda^0_1} - m_K - m_N}
  \nonumber\\
  &&+ \frac{1}{8\pi}\, \frac{1}{m_K + m_N}\,\Big[
  \frac{1}{\sqrt{3}}\, (3 - 2\alpha_2)\, g_{\pi NN_2}\Big] 
  \Big[\frac{2}{\sqrt{3}}\, \alpha_2\, g_{\pi NN_2}\Big] 
  \frac{1}{2\sqrt{m_{\Sigma}m_N}}\, \frac{1}{m_{\Lambda^0_2} - m_K - m_N}
  \nonumber\\
  &&-\frac{1}{8\pi}\, \frac{1}{m_K + m_N} 
  \Big[(2\alpha - 1)\, g_{\pi NN}\Big] 
  \Big[ 2\,(1 - \alpha)\, g_{\pi NN}\Big] 
  \frac{1}{2\sqrt{m_{\Sigma}m_N}}\, \frac{1}{m_{\Sigma^0} - m_K - m_N}
  \nonumber\\
  &&- \frac{1}{8\pi}\, \frac{1}{m_K + m_N} 
  \Big[(2\alpha_1 - 1)\, g_{\pi NN_1}\Big] 
  \Big[ 2\, (1 - \alpha_1)\, g_{\pi NN_1}\Big] 
  \frac{1}{2\sqrt{m_{\Sigma}m_N}}\, \frac{1}{m_{\Sigma^0_1} - m_K - m_N}
  \nonumber\\
  &&-\frac{1}{8\pi} \,\frac{1}{m_K + m_N} 
  \Big[(2\alpha_2 - 1)\, g_{\pi NN_2}\Big] 
  \Big[ 2\, (1 - \alpha_2)\, g_{\pi NN_2}\Big] 
  \frac{1}{2\sqrt{m_{\Sigma}m_N}}\, \frac{1}{m_{\Sigma^0_2} - m_K - m_N}
  \nonumber\\
  && = (- 0.015 + 0.006 - 0.001 + 0.005 - 0.001 + 0.002)\,{\rm m^{-3}_{\pi}} = 
  - 0.004\,{\rm m^{-3}_{\pi}}.
\end{eqnarray}
and 
\begin{eqnarray}\label{label4.5}
  (a^{\Sigma^-\pi^+}_{3/2})_R = - (a^{\Sigma^+\pi^-}_{3/2})_R = +\,0.082\,{\rm m^{-3}_{\pi}}.
\end{eqnarray}
The total P--wave scattering length of the reaction $K^-p \to
\Sigma^-\pi^+$ is equal to
\begin{eqnarray}\label{label4.6}
  2\,a^{\Sigma^-\pi^+}_{3/2} + a^{\Sigma^-\pi^+}_{1/2} = (2\,a^{\Sigma^-\pi^+}_{3/2} + a^{\Sigma^-\pi^+}_{1/2})_{SKT} + 0.160\,{\rm m^{-3}_{\pi}}.
\end{eqnarray}

\subsection{P--wave scattering lengths of  $2 a^{\Sigma^0\pi^0}_{3/2} + 
  a^{\Sigma^0\pi^0}_{1/2}$of inelastic channel $K^-p \to
  \Sigma^0\pi^0$}

The resonant parts of the  P--wave scattering lengths
$a^{\Sigma^0\pi^0}_{1/2}$ and $a^{\Sigma^0\pi^0}_{3/2}$ of the
reaction $K^-p \to \Sigma^0\pi^0$ are equal to
\begin{eqnarray}\label{label4.7}
  && \sum_R(a^{\Sigma^0\pi^0}_{1/2})_R =\nonumber\\
  &&=\frac{1}{8\pi} \,\frac{1}{m_K + m_N}\, 
  \Big[
  \frac{1}{\sqrt{3}}\, (3 - 2\alpha)\, g_{\pi NN}\Big] 
  \Big[\frac{2}{\sqrt{3}}\, \alpha\, g_{\pi NN}\Big] 
  \frac{1}{2\sqrt{m_{\Sigma}m_N}} \,\frac{1}{m_{\Lambda^0} - m_K - m_N}
  \nonumber\\
  &&+ \frac{1}{8\pi} \,\frac{1}{m_K + m_N}\,\Big[
  \frac{1}{\sqrt{3}}\, (3 - 2\alpha_1)\, g_{\pi NN_1}\Big] 
  \Big[\frac{2}{\sqrt{3}}\, \alpha_1 \,g_{\pi NN_1}\Big] 
  \frac{1}{2\sqrt{m_{\Sigma}m_N}}\, \frac{1}{m_{\Lambda^0_1} - m_K - m_N}
  \nonumber\\
  &&+ \frac{1}{8\pi}\, \frac{1}{m_K + m_N}\,\Big[
  \frac{1}{\sqrt{3}}\, (3 - 2\alpha_2)\, g_{\pi NN_2}\Big] 
  \Big[\frac{2}{\sqrt{3}}\, \alpha_2\, g_{\pi NN_2}\Big] 
  \frac{1}{2\sqrt{m_{\Sigma}m_N}}\, \frac{1}{m_{\Lambda^0_2} - m_K - m_N}
  \nonumber\\
  && = (- 0.015 + 0.006 - 0.001)\,{\rm m^{-3}_{\pi}} = 
- 0.010\,{\rm m^{-3}_{\pi}}.
\end{eqnarray}
and 
\begin{eqnarray}\label{label4.8}
  (a^{\Sigma^0\pi^0}_{3/2})_R = 0.
\end{eqnarray}
The total P--wave scattering length of the reaction $K^-p \to
\Sigma^0\pi^0$ is equal to
\begin{eqnarray}\label{label4.9}
  2\,a^{\Sigma^0\pi^0}_{3/2} + a^{\Sigma^0\pi^0}_{1/2} = (2\,a^{\Sigma^0\pi^0}_{3/2}
 + a^{\Sigma^0\pi^0}_{1/2})_{SKT} - 0.010\,{\rm m^{-3}_{\pi}}.
\end{eqnarray}

\subsection{P--wave scattering lengths of $2 a^{\Lambda^0\pi^0}_{3/2}
  + a^{\Lambda^0\pi^0}_{1/2}$of inelastic channel $K^-p \to
  \Lambda^0\pi^0$}

The resonant parts of the  P--wave scattering lengths
$a^{\Lambda^0\pi^0}_{1/2}$ and $a^{\Lambda^0\pi^0}_{3/2}$ of the
reaction $K^-p \to \Lambda^0\pi^0$ are equal to
\begin{eqnarray}\label{label4.10}
  && \sum_R(a^{\Lambda^0\pi^0}_{1/2})_R =\nonumber\\
  &&= \frac{1}{8\pi} \,\frac{1}{m_K + m_N}\, 
  \Big[-(2\alpha - 1)\, g_{\pi NN}\Big] 
  \Big[\frac{2}{\sqrt{3}}\, \alpha\, g_{\pi NN}\Big] 
  \frac{1}{2\sqrt{m_{\Lambda^0}m_N}} \,\frac{1}{m_{\Sigma^0} - m_K - m_N}
  \nonumber\\
  && + \frac{1}{8\pi} \,\frac{1}{m_K + m_N}\, 
  \Big[-(2\alpha_1 - 1)\, g_{\pi NN_1}\Big] 
  \Big[\frac{2}{\sqrt{3}}\, \alpha_1\, g_{\pi NN_1}\Big] 
  \frac{1}{2\sqrt{m_{\Lambda^0}m_N}} \,\frac{1}{m_{\Sigma^0_1} - m_K - m_N}
  \nonumber\\
  && + \frac{1}{8\pi} \,\frac{1}{m_K + m_N}\, 
  \Big[-(2\alpha_2 - 1)\, g_{\pi NN_2}\Big] 
  \Big[\frac{2}{\sqrt{3}}\, \alpha_2\, g_{\pi NN_2}\Big] 
  \frac{1}{2\sqrt{m_{\Lambda^0}m_N}} \,\frac{1}{m_{\Sigma^0_2} - m_K - m_N}
  \nonumber\\
  && = (0.006 - 0.005 - 0.001)\,{\rm m^{-3}_{\pi}} = 0.
\end{eqnarray}
and 
\begin{eqnarray}\label{label4.11}
  (a^{\Lambda^0\pi^0}_{3/2})_R &=&\frac{\sqrt{3} g^2_{\pi NN}}{36\pi m_N}
  \frac{1}{m_K + m_N}\,\frac{1}{m_{\Sigma^0_3} - m_N - m_K}\,\sqrt{\frac{m_{\Lambda^0}}{m_N}}\,\Big(1 + \frac{1}{4}\,\frac{m_K}{m_N}\,\frac{m_K + m_N}{m_{\Sigma^0_3}}\Big) =\nonumber\\
  &=&-\,0.137\,{\rm m^{-3}_{\pi}}.
\end{eqnarray}
The total P--wave scattering length of the reaction $K^-p \to
\Lambda^0\pi^0$ is equal to
\begin{eqnarray}\label{label4.12}
  2\,a^{\Lambda^0\pi^0}_{3/2} + a^{\Lambda^0\pi^0}_{1/2} = (2\,a^{\Lambda^0\pi^0}_{3/2}
 + a^{\Lambda^0\pi^0}_{1/2})_{SKT} - 0.274\,{\rm m^{-3}_{\pi}}.
\end{eqnarray}
\subsection{P--wave scattering lengths of inelastic reactions 
$K^-p \to Y\pi$ and energy level width of $np$ excited state of kaonic hydrogen}

According to the estimate Eq.(\ref{label3.27}), the contribution of
the P--wave scattering lengths $(2\,a^{Y\pi}_{3/2} +
a^{Y\pi}_{1/2})_{SKT}$ can be neglected in comparison with the
contribution of the baryon resonances. Therefore, below we neglect
$(2\,a^{Y\pi}_{3/2} + a^{Y\pi}_{1/2})_{SKT}$ for the estimate the
energy level width of the $np$ excited state of kaonic hydrogen.

Using Eqs.(\ref{label4.3}), (\ref{label4.6}) and (\ref{label4.12}) and
substituting them into Eq.(\ref{label2.13}), we compute the energy
level width of the $np$ excited state of kaonic hydrogen
\begin{eqnarray}\label{label4.13}
  \Gamma_{np} = \frac{32}{3}\,\frac{1}{n^3}\,\Big(1 - \frac{1}{n^2}\Big)\,\Gamma_{2p},
\end{eqnarray}
The partial width $\Gamma_{2p}$ of the energy level of the $2p$
excited state of kaonic hydrogen is equal to 
\begin{eqnarray}\label{label4.14}
\Gamma_{2p} = \frac{\alpha^5}{24}\,
  \Big(\frac{m_Km_N}{m_K + m_N}\Big)^4
  \sum_{Y\pi}(2\,a^{Y\pi}_{3/2} +
  a^{Y\pi}_{1/2})^2\,k^3_{Y\pi} = 2\,{\rm meV}
\end{eqnarray}
or $\Gamma_{2p} = 3\times 10^{12}\,{\rm sec^{-1}}$.  

The lifetime of the $2p$ state of kaonic hydrogen, defined by the
decays of kaonic hydrogen into hadronic states $(K^-p)_{2p} \to Y\pi$,
where $Y\pi = \Sigma^+\pi^-, \Sigma^-\pi^+, \Sigma^0\pi^0$ and
$\Lambda^0\pi^0$, is equal to $\tau_{2p} = 3.4\times 10^{-13}\,{\rm
  sec}$. It is much smaller than the lifetime of the $K^-$--meson,
$\tau_{K^-} = 1.24\times 10^{-8}\,{\rm sec}$ \cite{PDG00}, which is
the upper limit on the lifetime of kaonic hydrogen.  Thus, the rates
of the hadronic decays of kaonic hydrogen in the $np$ excited states
are comparable with the rates of the de--excitation of kaonic
hydrogen $np \to 1s$, caused by the emission of the $X$--rays
\cite{XR1}--\cite{XR7}.

The result obtained for the partial width of
the excited $2p$ state of kaonic hydrogen, given by
Eq.(\ref{label4.14}), is important for the theoretical analysis of the
$X$--ray yields in kaonic hydrogen \cite{XR1}--\cite{XR7}.

\section{Conclusion} 
\setcounter{equation}{0}

The quantum field theoretic model of the description of low--energy
$\bar{K}N$ interaction in the S--wave state near threshold, which we
have suggested in \cite{IV3,IV4}, is extended on the analysis of
low--energy $\bar{K}N$ interactions in the P--wave state near
threshold. We would like to emphasize that our approach to the
description of low--energy $\bar{K}N$ interaction in the S--wave state
near threshold agrees well with the non--relativistic Effective Field
Theory based on ChPT by Gasser and Leutwyler, which has been recently
applied by Mei\ss ner {\it et al.} \cite{UM04} to the calculation of
the energy level displacement of the $ns$ state of kaonic hydrogen and
systematic corrections to the energy level displacement of the $ns$
state, caused by QCD isospin--breaking and electromagnetic
interactions. The result for the energy level displacement of the $ns$
state of kaonic hydrogen has been obtained in \cite{UM04} in terms of
the S--wave scattering lengths $a^0_0$ and $a^1_0$ of $\bar{K}N$
scattering with isospin $I = 0$ and $I = 1$, respectively. The S--wave
scattering lengths $a^0_0$ and $a^1_0$ have been treated as free
parameters of the approach. Using our results for the S--wave
scattering lengths $a^0_0$ and $a^1_0$ \cite{IV3,IV4} and keeping
leading terms in QCD isospin--breaking and electromagnetic
interactions, i.e. accounting for only the contribution of Coulombic
photons, we have shown that the numerical value of the energy level
displacement of the $ns$ state of kaonic hydrogen, computed by Mei\ss
ner {\it et al.} \cite{UM04}, agrees well with both our theoretical
prediction \cite{IV3} and recent experimental data by the DEAR
Collaboration \cite{DEAR1} within 1.5 standard deviations for the
shift and one standard deviation for the width. Hence, our approach
to the description low--energy dynamics of strong low--energy
$\bar{K}N$ interactions at threshold agrees well with general
description of strong low--energy interactions of hadrons within
non--relativistic Effective Field Theory based on ChPT
\cite{UM04}--\cite{JG99}.

The detection of the $X$--rays of the $X$--ray cascade processes,
leading to the de--excitation of kaonic hydrogen from the excited
states to the ground state, is the main experimental tool for the
measurement of the energy level displacement of the ground state of
kaonic hydrogen, caused by strong low--energy interactions
\cite{DEAR1,DEAR3}. The main transitions in kaonic hydrogen, which are
measured experimentally for the extraction of the energy level
displacement of the ground state, are  $3p \to 1s$ and $2p \to 1s$,
i.e.  the reactions $(K^-p)_{3p} \to (K^-p)_{1s} + \gamma$ and
$(K^-p)_{2p} \to (K^-p)_{1s} + \gamma$.

As has been pointed out by Markushin and Jensen the yields of
$X$--rays of these transitions are quite sensitive to the value of
$\Gamma_{2p}$ \cite{XR7}.  Using $\Gamma_{2p}$ as an input parameter
taking values from the region $0.1\,{\rm meV} \le \Gamma_{2p} \le
0.9\,{\rm meV}$, Markushin and Jensen \cite{XR7} have found that their
theoretical predictions for the $X$--ray yields in kaonic hydrogen
agree well with the experimental data on the $X$--ray yields detected
by the KEK Collaboration \cite{KEK1}, which have been used for the
extraction of the energy level displacement of the ground state of
kaonic hydrogen, for $\Gamma_{2p} = 0.3\,{\rm meV} = 4.6\times
10^{11}\,{\rm sec^{-1}}$ and $\epsilon_{1s} = 320\,{\rm eV}$ and
$\Gamma_{1s} = 400\,{\rm eV}$.

Recent experimental data on the energy level displacement of the
ground of kaonic hydrogen obtained by the DEAR Collaboration
\cite{DEAR1} by a factor of 2 smaller than the experimental data by
the KEK Collaboration \cite{KEK1}. Our theoretical analysis of the
energy level displacement of the $2p$ excited state of kaonic hydrogen
has shown that the rate of the hadronic decays of kaonic hydrogen from
the $2p$ excited state is equal to $\Gamma_{2p} = 2\,{\rm meV} =
3\times 10^{12}\,{\rm sec^{-1}}$, which is an order of magnitude
larger than the phenomenological value $\Gamma_{2p} =
0.3\,{\rm meV} = 4.6\times 10^{11}\,{\rm sec^{-1}}$, used by Markushin
and Jensen as an input parameter \cite{XR7}.

Thus, the computed value $\Gamma_{2p} = 2\,{\rm meV}$ of the energy
level of the $2p$ excited state of kaonic hydrogen can be applied to
the theoretical analysis of the $X$--ray yields in kaonic hydrogen of
recent experimental data by the DEAR Collaboration \cite{DEAR1} using
the the following input parameters: 1) the experimental setup
\cite{DEAR3} and 2) the theoretical predictions for the hadronic
energy level displacements of the $2p$ state, $\epsilon_{2p} = -\,0.6\,
{\rm meV}$, $\Gamma_{2p} = 2.0\, {\rm meV}$, and the ground state,
$\epsilon_{1s} = 203\, {\rm eV}$ and $\Gamma_{1s} = 226\, {\rm eV}$,
of kaonic hydrogen \cite{IV3}.

\section{Comment on the result}

After the manuscript has been posted at archive Faifman and Men'shikov
have presented the calculated yields for the $K$--series of $X$--rays
for kaonic hydrogen in dependence of the hydrogen density \cite{MF05}.
They have shown that the use of the theoretical value $\Gamma_{2p} =
2\,{\rm meV}$ of the width of the $2p$ state of kaonic hydrogen,
computed in our work, leads to good agreement with the experimental
data, measured for the $K_{\alpha}$--line by the KEK Collaboration
\cite{KEK1}.  They have also shown that the results of cascade
calculations with other values of the width of the $2p$ excited state
of kaonic hydrogen, used as an input parameter, disagree with the
available experimental data. The results obtained by Faifman and
Men'shikov contradict to those by Jensen and Markushin \cite{XR7}.
Therefore, as has been accentuated by Faifman and Men'shikov
\cite{MF05}, the further analysis of the experimental data by the DEAR
Collaboration should allow to perform a more detailed comparison of
the theoretical value $\Gamma_{2p} = 2\,{\rm meV}$ with other
phenomenological values of the width of the $2p$ state of kaonic
hydrogen $\Gamma_{2p}$, used as input parameters.


\begin{thebibliography}{9}
\bibitem{IV3}
A. N. Ivanov, M. Cargnelli, M. Faber, J. Marton, N. I. Troitskaya, and 
J. Zmeskal,
Eur. Phys. J. A {\bf 21}, 11 (2004); nucl--th/0310081.
\bibitem{DEAR1} 
M. Cargnelli {\it et al.} (DEAR Collaboration),
Kaonic Nuclear Clusters -- Miniworkshop, (IMEP, Wien), 9 February 2004;
M. Cargnelli {\it et al.} (DEAR Collaboration),
in Proceedings of HadAtom03 Workshop, 13--17 October 2003,
ECT$^*$ (Trento Italy), hep--ph/0401204.
\bibitem{UM04}
U.--G. Mei\ss ner, U. Raha, and A. Rusetsky,
Eur. Phys. J. C {\bf 35}, 349 (2004); hep--ph/0402261.
\bibitem{JG83}
J. Gasser and H. Leutwyler, 
Phys. Lett. B {\bf 125}, 321, 325 (1983).
\bibitem{JG99}
J. Gasser, 
Nucl. Phys. Proc. Suppl. {\bf 86}, 257 (2000) and references therein.
H. Leutwyler, PiN Newslett. {\bf 15}, 1 (1999);
Ulf-G. Mei\ss ner, 
PiN Newslett. {\bf 13}, 7 (1997);
H. Leutwyler,
 Ann. of Phys. {\bf 235},
165 (1994);
G. Ecker, 
Prog. Part. Nucl. Phys. {\bf 36}, 71 (1996); 
Prog. Part. Nucl. Phys. {\bf 35}, 1 (1995); 
Nucl. Phys. Proc. Suppl. {\bf 16}, 581 (1990);
J. Gasser, Nucl. Phys. B {\bf 279}, 65 (1987);
 J. Gasser, H. Leutwyler, 
Nucl. Phys. B {\bf 250}, 465 (1985); 
Ann. of  Phys. {\bf 158}, 142 (1984);
Phys. Lett. B {\bf 125}, 321 (1983).
\bibitem{IV4}
A. N. Ivanov, M. Cargnelli, M. Faber, H. Fuhrmann, V. A. Ivanova, J. Marton, 
N. I. Troitskaya, and J. Zmeskal, 
Eur. Phys. J. A {\bf 23}, 79 (2005), nucl--th/0406053.
\bibitem{XR1}
T. B. Day, G. A. Snow, and J. Sucher,
Phys. Rev. Lett. {\bf 3}, 61 (1959);
R. K. Adair,
Phys. Rev. Lett. {\bf 3}, 438 (1959);
M. Leon and H. A. Bethe,
Phys. Rev. {\bf 127}, 636 (1962).
\bibitem{XR2}
T. E. O. Ericson and F. Scheck,
Nucl. Phys. B {\bf 19}, 450 (1970).
\bibitem{XR3}
E. Borie and M. Leon,
Phys. Rev. A {\bf 21}, 1460 (1980).
\bibitem{XR4}
T. Koike, T. Harada, and Y. Akaishi,
Phys. Rev. C {\bf 53}, 79 (1996).
\bibitem{XR5}
T. P. Terada and R. S. Hayano,
Phys. Rev. C {\bf 55}, 73 (1997).
\bibitem{XR6}
M. P. Faifman {\it et al.},
Frascati Physics Series Vol. XVI,pp. 637--641,
{\it PHYSICS AND DETECTORS FOR DA$\Phi$NE}--Frascati, Nov.16--19, 1999,
{\it Physics of the Atomic Cascades in Kaonic and Hydrogen and Deuterium};
M. P. Faifman and L. I. Men'shikov,
{\it Cascade Processes in Kaonic and Muonic Atoms},
Proceeding of International Workshop on {\it EXOTIC ATOMS -- FUTURE 
PERSPECTIVES} at Institute of Medium Energy Physics of Austrian
Academy of Sciencies, November 28--30, 2002, Vienna, Austia,
pp.185--196.
\bibitem{XR7}
V. E. Markushin and T. S. Jensen,
Nucl. Phys. A {\bf 691}, 318c (2001);
T. S. Jensen and  V. E. Markushin,
Nucl. Phys. A {\bf 689}, 537 (2001);
Eur. Phys. J. D {\bf 19}, 165 (2002); 
{\it Collisional de--excitations of exotic hydrogen atoms in highly excited 
states.I. Cross sections.}, physics/0205076; 
{\it Collisional de--excitations of exotic hydrogen atoms in highly excited 
states.II. Cascade calculations.}, physics/0205077.
\bibitem{IV2} 
A. N. Ivanov, M. Faber, A. Hirtl, J. Marton, and N. I. Troitskaya,
Eur. Phys. J. A {\bf 19}, 413 (2004); nucl--th/0310027.
\bibitem{MA72}
{\it HANDBOOK OF MATHEMATICAL FUNCTIONS WITH Formulas, 
Graphs, and Mathematical Tables}, edited by M. Abramowitz 
and I. A. Stegun, National Bureau of Standards, Applied 
Mathematics Series $\,\bullet\,$ 55, 1972.
\bibitem{LL65}
L. D. Landau and E. M. Lifshitz,
in {\it QUANTUM MECHANICS}, Volume 3 of Course of Theoretical
Physics, Pergamon Press, Oxford, 1965, pp.116--128.
\bibitem{SG67}
S. Gasiorowicz,
in {\it ELEMENTARY PARTICLE PHYSICS},
John $\&$ Sons, Inc., New York, 1967.
\bibitem{MN79}
M. M. Nagels {\it et al.},
Nucl. Phys. B {\bf 147}, 189 (1979).
\bibitem{TE88} 
T. E. O. Ericson and W. Weise, 
in {\it PIONS AND NUCLEI}, Clarendon Press, Oxford, 1988.
\bibitem{PDG00}
D. E. Groom {\it et al.} (Particle Data Group),
Eur. Phys. J. C {\bf 15}, 1 (2000).
\bibitem{MR96}
C.--H. Lee, D.--P. Min, and M. Rho,
Phys. Lett. B {\bf 326}, 14 (1994);
C.--H. Lee, G. E. Brown, and M. Rho,
Phys. Lett. B {\bf 335}, 266 (1994);
C.--H. Lee, G. E. Brown, D.--P. Min, and M. Rho,
Nucl. Phys. A {\bf 585}, 401 (1995);
C.--H. Lee, D.--P. Min, and M. Rho,
Nucl. Phys. A {\bf 602}, 334 (1996).
\bibitem{SG69}
S. Gasiorowicz and D. A. Geffen,
Rev. Mod. Phys. {\bf 41}, 531 (1969).
\bibitem{SA68}
S. L. Adler and R. Dashen,
in {\it CURRENT ALGEBRAS}, Benjamin, New York 1968.
\bibitem{ME71}
M. Ericson and  M. Rho,
Phys. Lett. B {\bf 36}, 93 (1971)
\bibitem{HP73}
V. De Alfaro, S. Fubini, G. Furlan, and C. Rossetti,
in {\it CURRENTS IN HADRON PHYSICS},
North--Holland Publishing Co., Amsterdam $\,\bullet\,$ London,
American Elsevier Publishing Co., Inc.,
New York, 1973.
\bibitem{ER72}
E. Reya,
Phys. Rev. D {\bf 6}, 200 (1972);
Phys. Rev. D {\bf 7}, 3472 (1973);
Rev. Mod. Phys. {\bf 46}, 545 (1974).
\bibitem{MS77}
J. F. Gunion, P. C. Mcamee, and M. D. Scadron,
Nucl. Phys. B {\bf 123}, 445 (1977). 
\bibitem{BC79}
B. di Claudio, A. M. Rodriguez--Vargas, and G. Violini,
Z. Phys. C {\bf 3}, 75 (1979).
\bibitem{TE87}
T. E. O. Ericson,
Phys. Lett. B {\bf 195}, 116 (1987).
\bibitem{EK98}
E. E. Kolomeitsev,
in {\it KAONEN IN KERNMATERIE}, PhD, 1998;\\
http://www.physik.tu--dresden.de/publik/1998/
diss$_-$kolomeitsev.ps
\bibitem{RJ77}
R. L. Jaffe,
Phys. Rev. D {\bf 13}, 267, 281 (1977).
\bibitem{NA80}
N. N. Achasov, S. A. Devyanin and G. N. Shestakov,
Sov. J. Nucl. Phys. {\bf 32}, 566 (1980); 
Phys. Lett. {\bf B 96}, 168
(1980); Phys. Lett. {\bf B 108}, 134 (1982); 
Z. Phys. {\bf C 16}, 55 (1982);
Sov. Phys. Usp. {\bf 27}, 161 (1984);
N. N. Achasov and G. N. Shestakov,
Z. Phys. C {\bf 41}, 309 (1988);
N. N. Achasov, Nucl. Phys. B (Proc.Suppl.) {\bf 21}, 189 (1991);
N. N. Achasov and G. N. Shestakov,
Sov. Phys. Usp. {\bf 34}, 471 (1991);
N. N. Achasov, V. V. Gubin, and V. I. Shevchenko,
Phys. Rev. D {\bf 56}, 203 (1997);
N. N. Achasov and V. N. Gubin,
Phys. Rev. D {\bf 56}, 4084 (1997);
N. N. Achasov,
Phys. Usp. {\bf 41}, 1149 (1998), hep-ph/9904223;
N. N. Achasov and  G.N. Shestakov,
Phys. Atom. Nucl. {\bf 62}, 505 (1999);
N. N. Achasov,
Nucl. Phys. A {\bf 675}, 279c (2000)
N. N. Achasov and  A.V. Kiselev,
Phys. Lett. B {\bf 534}, 83 (2002);
N. N. Achasov,
Phys. Atom. Nucl. {\bf 65}, 546 (2002).
\bibitem{SK03}
S. Krewald, R. H. Lemmer, and F. P. Sassen,
{\it Lifetime of kaonium}, hep-ph/0307288;
F. P. Sassen, S. Krewald, and J. Speth,
Phys. Rev. D {\bf 68}, 036003 (2003);
S. Krewald,
{\it Kaonium and meson--exchange models of meson--meson interactions},
Invited talk at the Workshop on {\it HADATOM03} at ECT$^*$ in Trento, 
12--18 October 2003, Italy.
\bibitem{DAPHNE}
M. Primavera,
{\it Results from DAPHNE},
Proceedings of the Workshop 
on {\it CHIRAL DYNAMICS} at University of Bonn, 
Bonn 8--13 September 2003, Germany, p.22, hep--ph/0311212.
\bibitem{PSI2}
H.--Ch. Schr\"oder {\it et al.},
Eur. Phys. J. C {\bf 21}, 473 (2001).
\bibitem{TE03}
T. E. O. Ericson, B. Loiseau, and S. Wycech,
Phys. Lett. B {\bf 594}, 76 (2004).
\bibitem{AI99}
A. N. Ivanov, M. Nagy, and N. I. Troitskaya,
Phys. Rev. C {\bf 59}, 451 (1999).
\bibitem{ST1}
J. Gasser, H. Leutwyler, and  M. E. Sainio,
Phys. Lett. B {\bf 253}, 252, 260 (1991); 
V. Bernard, N. Kaiser, and Ulf--G. Mei\ss ner,
Z. Phys. C {\bf 60}, 111 (1993);
B. Borasoy,  
Eur. Phys. J. C {\bf 8}, 121 (1999);
J. Gasser and M. E. Sainio,
{\it SIGMA--TERM PHYSICS}, Invited talk given at the Workshop
{\it Physics and Detectors for {\rm DA$\Phi$NE}}, Frascati,
November 16--19, 1999.
\bibitem{DEAR3}
G. Beer {\it et al.},
{\it Kaonic Hydrogen: Status of the DEAR Experiment},
Progress of Theoretical Physics Supplement {\bf 149}, pp.240--246 (2003).
\bibitem{KEK1}
T. M. Ito {\it et al.} (KEK Collaboration),
Phys. Rev. C {\bf 58}, 2366 (1998).
\bibitem{MF05} 
M. P. Faifman and L. I. Men'shikov, 
{\it A new approach
    to kinetic analysis of cascade processes in $\mu^-p$ and
    $K^-p$ hydrogen atoms}, Invited talk at Workshop on exotic atoms
  ``EXA${^{05}}$'' at Stefan Meyer Institute of subatomic Physics 
of Austrian Academy of Sciences, 21--25 February 2005, Vienna, Austria;\\
http://www.oeaw.ac.at/smi/exa05/program.htm
\end{thebibliography}
\end{document}